\documentclass[11pt]{article}          
\usepackage{hyperref}
\usepackage{graphicx}
\usepackage{epsfig}
\usepackage{xspace}
\usepackage{array}

\begin{document}

\begin{center}

{\Large The effect of wavefront corrugations on fringe motion in
  an astronomical interferometer with spatial filters}

\vspace{1cm}

{\large Robert Tubbs}

\vspace{0.5cm}

{\large Leiden Observatory, P.O. Box 9513, 2300 RA Leiden, Netherlands}

\vspace{0.1cm}

  \includegraphics[width=3.2cm]{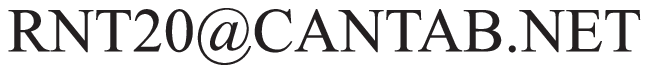}

\end{center}

\section*{Abstract}Numerical simulations of atmospheric turbulence and AO wavefront
correction are performed to investigate the timescale for fringe
motion in optical interferometers with spatial filters. These
simulations focus especially on partial AO correction, where only a
finite number of Zernike modes are compensated. The fringe motion is
found to depend strongly on both the aperture diameter and the level
of AO correction used. In all of the simulations the coherence
timescale for interference fringes is found to decrease dramatically
when the Strehl ratio provided by the AO correction is 
$\mathrel{\hbox to 0pt{\lower 3pt\hbox{$\mathchar"218$}}
     \raise 2.0pt\hbox{$\mathchar"13C$}\hss} 30\%$. For AO systems
which give perfect compensation of a limited number of Zernike modes,
the aperture size which gives the optimum signal for fringe phase
tracking is calculated. For AO systems which provide noisy
compensation of Zernike modes (but are perfectly piston-neutral), the
noise properties of the AO system determine the coherence timescale of
the fringes when the Strehl ratio is $\mathrel{\hbox to 0pt{\lower 3pt\hbox{$\mathchar"218$}}
     \raise 2.0pt\hbox{$\mathchar"13C$}\hss} 30\%$.

\section{Introduction}
The field of astronomical interferometry has matured over the past two
decades with a substantial number of interferometric arrays being
constructed around the world\cite{monnier03}. The majority of the
interferometric instruments at existing arrays utilise pupil-plane
Michelson beam combination. These instruments probe the structure on
much smaller angular scales than can be resolved by the individual
telescopes which make up the array, and the light which is useful for
an interferometric measurement typically resides within a region which
is unresolved by the individual telescopes (so the parts of an
astronomical source which are studied interferometrically typically
lie within a single diffraction-limited point spread function for one
of the telescopes making up the interferometric array). It is common
at such interferometers for the incoming wavefronts to be spatially
filtered by using a pinhole or a single-mode fibre to select light
from the central core of the Airy pattern in the image plane of each
telescope, and to exclude all light from the periphery of the
Point-Spread Function (PSF)\cite{keen01,monnier03}, as this light
would decrease the signal-to-noise ratio for measurements of the
fringe visibility.

Optical wavefronts from an astronomical source passing through the
Earth's atmosphere are perturbed by fluctuations in air density and
water vapour density\cite{roddier81} resulting in rotation of the
phase of the optical wavefronts. Experimental measurements indicate
that the spatial and temporal power spectra of these fluctuations are
usually well fit by \emph{Kolmogorov
models}\cite{tatarski61,kolmogorov41a,kolmogorov41b} for spatial
scales between the inner and outer scales of turbulence. 

If there are atmospherically-induced wavefront perturbations across
one of the telescope apertures in an interferometric array there may
be a significant reduction in the on-axis Strehl ratio measured at
that telescope, with only a fraction of the light falling into an
on-axis Airy disk. This results in a reduction in the flux entering a
spatial filter positioned on the optical axis, as well as fluctuations
in the optical phase of that light. The flux reduction has been
studied elsewhere\cite{fusco04, monnier03}, and so I concentrate here
on the fluctuations in the optical phase found after the
spatial-filtering process.

For an optical interferometer formed from two telescopes of finite
diameter observing a source which is unresolved by the individual
telescopes, the contribution of atmospheric turbulence can be split
into two components:
\begin{enumerate}
\item The difference in the mean optical path for the two different
      light paths to the star through the two apertures -- the
      differential piston mode\cite{conan95}
      Zernike\cite{noll76} component
\item The wavefront corrugations over each individual telescope
      aperture of the interferometer (the higher order Zernike modes)
\end{enumerate}

If each telescope has a spatial filter, then both of the components
listed above will affect the timescales over which the fringes move
(and hence those over which fringe measurements must be made). The
contribution from component 2 results from the strong coupling of the
high-order Zernike modes for wavefronts in the telescope aperture
plane to the optical phase output from the spatial filter (as is
customary, I use the term ``optical phase'' to indicate the phase of
the monochromatic optical wavefronts measured relative to the
wavefront phase before the light was perturbed by the atmosphere or
optical instrumentation). An adequate analytical model for the
contribution of this to the timescale for fringe motion in an
interferometer has not yet been developed. This has forced most
previous authors either to limit their discussions of fringe motion to
small apertures (e.g. Ref.~\cite{buscher88}) or to ignore the
effect of term 2 above even for large
apertures\cite{conan95,darcio99}, which would be a valid assumption
for the case of ideal adaptive optics (AO) correction providing Strehl
ratios of $100\%$.

The numerical simulations presented here address the timescale for
fringe motion in an interferometer with large apertures and only
partial AO correction. Although simulations such as these have been
performed previously when optimising the design of interferometers, as
far as I am aware detailed results have not been published in the
literature. In order to keep the results as general as possible, the
atmosphere was modelled using unchanging Kolmogorov phase screens
blown past in a wind-scatter model\cite{roddier82}, and the AO
correction in the simulations was restricted to compensation of a
finite number of Zernike modes.

\section{Simulation method}
\label{sect:sim_method}

Simulations were developed to investigate the motion of interference
fringes in a two-element optical interferometer. The optical modelling
was limited to one pupil plane and one image plane and the wavefront
was sampled as complex numbers in a rectangular grid of points fine
enough to avoid aliasing errors. Transformations from the pupil plane
to the image plane were performed by applying an FFT to the grid of
points. Input wavefront perturbations were provided by two Taylor
screens introducing Kolmogorov phase perturbations with a large outer
scale ($>1000 r_{0}$ -- further details of the simulations can be
found in Ref.~\cite{tubbs04a}). Both screens moved at the same
speed but at an angle of $120^{\circ}$ with respect to each other
(chosen arbitrarily). This arrangement provided a speckle coherence
timescale $\Delta_{\tau}$ (see Ref.~\cite{roddier82}) which was
essentially independent of aperture diameter. The results are not
expected to depend critically on the dispersion in the wind velocities
(the wind dispersion only strongly effects the speckle pattern
properties in telescopes without AO\cite{roddier82}), but for
realistic simulations of existing interferometers researchers are
encouraged to select wind velocity profiles which match the conditions
above their interferometer as closely as possible. The time unit in
the simulations was defined as the time taken for each phase screen to
move by the coherence length $r_{0}$ for the full atmosphere. The
simulations were repeated at discrete time-steps separated by $1/10$th
of a time unit.

The wavefront perturbations introduced by the atmosphere were
calculated at a grid of points at fixed locations in the telescope
pupil plane, with the values being based on linear interpolation
between the grid points making up the moving Taylor screen arrays to
minimise aliasing effects.

Experience from previous simulations\cite{tubbs04a} and from real
interferometers indicates that the fringe motion in an interferometer
is only affected significantly by wavefront corrugations on the
individual telescopes when the image breaks up into multiple
speckles. Each speckle in a multi-speckle image has a random
(wavelength-dependent) phase rotation relative to the mean piston
phase of the wavefronts in the pupil plane. Adjustment of Zernike
modes above the piston mode (e.g. AO correction) causes changes to the
image plane speckle pattern, resulting in changes to the phases
measured in the image plane even if the pupil-plane piston mode is
unaffected. The simulations presented here will investigate the
effects of the residual wavefront errors after partial AO correction
on the phases measured in an image-plane spatial filter.

The AO correction of the wavefront perturbations was simulated by
numerically correcting the lowest Zernike modes in the pupil plane
(excluding piston). Care was taken to ensure that the pupil plane
piston mode component was not changed during this correction
process. Eight levels of AO correction were simulated as listed in
Table~\ref{table:ao_list}.

\begin{table}[h]
{\bf \caption{Models for AO wavefront correction\label{table:ao_list}}}\begin{center}
\begin{tabular}{lp{3.5cm}p{1.7cm}}\hline
Model & Highest Zernike mode corrected ($j$) & Diameter $D_{j}$$^{\dag}$ \\ \hline
$1$ & $3$ & $3.3r_{0}$ \\ 
$2$ & $6$ & $5.2r_{0}$ \\ 
$3$ & $10$ & $6.9r_{0}$ \\ 
$4$ & $15$ & $8.6r_{0}$ \\ 
$5$ & $21$ & $10r_{0}$ \\ 
$6$ & $28$ & $12r_{0}$ \\ 
$7$ & $36$ & $13r_{0}$ \\ 
$8$ & $45$ & $15r_{0}$ \\ \hline
\end{tabular}
\end{center}
\begin{center}
\parbox{7cm}
{
\begin{tabular}{lp{6.2cm}}
$^{\dag}$ & The aperture diameter where the variance of the wavefront phase
would be $1$~rad$^2$
\end{tabular}
}
\end{center}
\end{table}

The long exposure on-axis Strehl ratio $R$ is described by\cite{fusco04}:

\begin{equation}
R \ge \exp(-\sigma^{2})
\end{equation}

where $\sigma^{2}$ is the residual variance in the wavefront phase in
the on-axis direction. If $\sigma^{2}<1$ rad$^{2}$ after AO
correction, the point-spread function for the telescope will be almost
diffraction-limited with a Strehl ratio of more than $37\%$, and the
measured optical phase after spatial filtering is expected to closely
match the mean piston-mode component. A minimum diameter $D_{j}$ was
defined as that where the residual phase variance $\sigma^{2}$ was
$1$~rad$^{2}$, given by:

\begin{equation}
D_{j}/r_{0}=\left(a_{j}\right )^{-3/5}
\end{equation}

where $a_{j}$ is the numerical coefficient in the $j$th
Zernike-Kolmogorov residual error formula of Table IV from
Ref.~\cite{noll76}. Values of $D_{j}$ are included in
Table~\ref{table:ao_list} for each of the models. Based on these
values, a range of aperture diameters between $4r_{0}$ and $20r_{0}$
were chosen for the all simulations presented here.

In order to reduce the complexity of the simulations, only one arm of
the interferometer was modelled. In an interferometer with
widely-separated apertures the wavefront corrugations measured at the
two apertures will be uncorrelated, and the RMS fringe phase change in
the interferometer will be $\sqrt{2}$ times greater than the RMS
change in optical phase measured at either one of the telescopes. The
results calculated here for a single telescope can thus be readily
extrapolated to the case of a long-baseline interferometer.

After AO correction, the optical wavefronts were spatially filtered by
summing over a uniform circular disk in the telescope image plane. The
circular disk used had a diameter corresponding to an angle of
$\lambda/D$ on the sky (typical for pinhole or single-mode fibre
spatial filters\cite{keen01}), where $\lambda$ was the mean wavelength
for the observation and $D$ was the aperture diameter used. This
summed flux was used for the remainder of the simulation.

In an optical interferometer operating with a finite optical bandpass,
interference fringes are only found over a narrow range of optical
path difference (a narrow range of differential optical delay between
the two telescopes and the beam combiner). The maximum level of
interference is found when the optical delay (optical path) along one
arm of the interferometer is adjusted so that the gradient of optical
phase difference with wavelength is at a minimum. This delay setting
occurs when the reciprocal of the group velocity integrated along the
both optical paths from the star is equal, and is called the group
delay.

The optical phase delay was tracked in the simulations using a
three-stage process designed to emulate what might be done at a real
interferometer:
\begin{enumerate}
\item The phase difference between light in two closely-spaced
  wavelength channels was used to calculate a group delay.
\item A change in the optical delay in one interferometer arm was
  simulated by rotating the phases in each channel, subtracting the
  group delay offset measured in 1. (using the $\frac{2\pi}{\lambda}$
  factor between phase and optical path to correctly take account of
  the wavelength $\lambda$ of each channel).
\item An average phase was calculated for the two
  channels (averaging vector representations of the electric
  field to avoid phase-wrapping errors).
\item The group delay offset subtracted in 2. was added back to give
  an unwrapped phase (corresponding to $2\pi$ times the optical path
  offset measured in wavelengths).
\end{enumerate}
In these simulations the two wavelengths used were separated by $1\%$
of the mean wavelength, a sufficiently narrow bandpass that the effect
of the variation in the light path through the atmosphere with
wavelength due to atmospheric refraction can be ignored. Care was
taken to ensure that the same sampling grid was used in the image
plane for both wavelengths, that the pupil plane representations of
the wavefronts were truncated at the same aperture diameter, and that
the wavelength difference was correctly taken into account when
interpolating the wavefront perturbations from the atmospheric Taylor
screens.

\section{Results}
\label{sect:results}

\begin{figure}
  \centering
  \includegraphics[width=8.3cm]{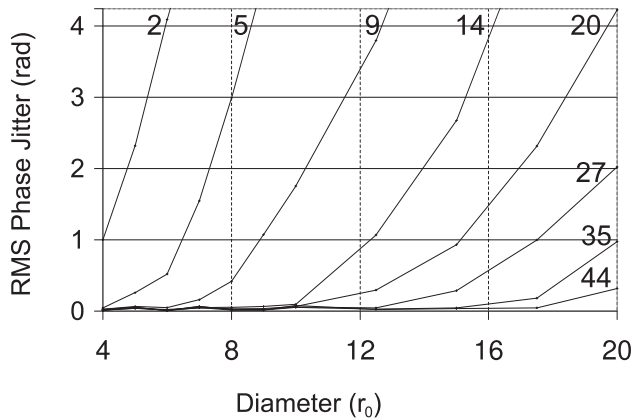}
  \caption{The RMS jitter in the optical phase in an interferometer
  with spatial filters, plotted against aperture diameter. The jitter
  results from wavefront corrugations in the telescope aperture
  plane. Each line connects models with a given level of AO
  correction, and is labelled with the number of Zernike modes which
  have been suppressed. The phase jitter plotted represents the RMS
  difference between the optical phase calculated for that model and
  the phase delay after ideal AO correction.}
  \label{fig:phase_jitter}
  \end{figure}

In order to separate the effects of higher Zernike modes from the
piston mode across the telescope aperture, the piston mode measured in
the pupil plane was subtracted from the phases measured at the spatial
filter for each of the simulations. The RMS of the result is a measure
of the phase jitter introduced by wavefront corrugations across the
simulated telescope apertures. The RMS jitter in a two-element
interferometer is $\sqrt{2}$ times larger than the jitter in the
unwrapped phase for a single telescope, and is shown for simulations
of $8192$ time-steps in Fig.~\ref{fig:phase_jitter}.

Phase jitter in the interference fringes will significantly degrade
the interferometric performance when it reaches $\sim 1$~rad
RMS. In these simulations this level of phase jitter is only reached
if the aperture diameter is significantly greater than $D_{j}$ (see
Table~\ref{table:ao_list}), corresponding to AO Strehl ratios
significantly less than $37\%$. It is clear, however, that the phase
jitter does become significant with large telescope apertures and
non-ideal AO correction, and should thus be taken into account in
signal-to-noise calculations for interferometer performance.

\begin{figure}
  \centering
  \includegraphics[width=8.3cm]{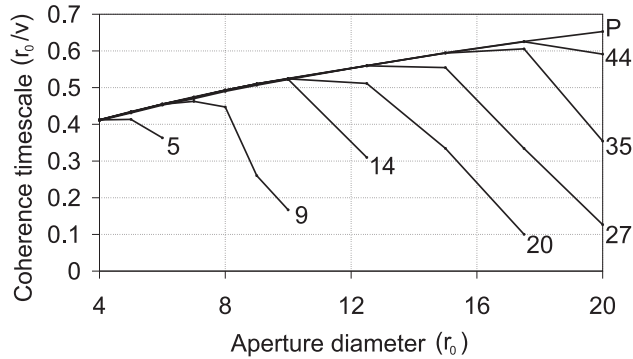}
  \caption{The coherence timescale for the optical phase for a range
  of aperture diameters and different levels of AO correction. The
  labels on the lines indicate the number of Zernike modes corrected
  from tip-tilt upwards. The curve labelled ``P'' corresponds to the
  coherence timescale of the piston mode, indicating the result
  expected for ideal AO correction. The endpoint of each curve is
  determined by the time-resolution of the simulations performed.}
  \label{fig:timescales}
  \end{figure}

If the phase jitter is $\mathrel{\hbox to 0pt{\lower
3pt\hbox{$\mathchar"218$}} \raise 2.0pt\hbox{$\mathchar"13E$}\hss}
1$~rad RMS, then the temporal frequency spectrum of the phase
jitter will affect the optimum exposure time for fringe observations
in the interferometer. Temporal structure functions of the total
optical phase (including both the phase jitter and the piston mode)
were calculated for each model of AO wavefront correction and each
aperture size. From these structure functions the timescale for the
phase to change by $1$~rad in the interferometer was calculated
using linear interpolation. This corresponds to the coherence
timescale for the fringe phase in an optical interferometer having
this level of AO correction. The calculated timescales are shown in
Fig.~\ref{fig:timescales} for all aperture sizes where the timescale
was sufficiently long to allow meaningful interpolation from the
sampled structure function.

I will now look at how the signal-to-noise in an interferometer can be
optimised, allowing the usage of observing time to be minimised. I
will assume here that there are limits on the instrumentation
available (so that, for example, the level of AO correction cannot be
arbitrarily increased in order to maximise the signal-to-noise ratio),
and that the seeing conditions are fixed. A commonly used approach to
improving the signal-to-noise ratio under these constraints at
astronomical interferometers is to adjust the telescope aperture
diameter (e.g. the tip-tilt corrected COAST apertures were stopped
down to 14cm for the first interferometric imaging
observations\cite{baldwin96}). In order to maintain the same level of
AO correction with a reduced aperture diameter, the AO system must
clearly be optically matched to the new aperture size.

\begin{figure}
  \centering
  \includegraphics[width=8.3cm]{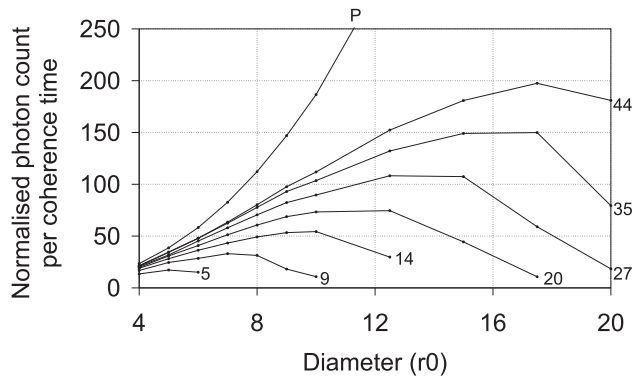}
  \caption{The photon count through the spatial filter per coherence
  timescale normalised so that a diffraction-limited aperture of
  diameter $r_{0}$ would give unity. The labels on the lines indicate
  the number of Zernike modes corrected from tip-tilt upwards. The
  curve labelled ``P'' corresponds to the result expected for ideal AO
  correction.}
  \label{fig:flux}
  \end{figure}

The signal-to-noise ratio for tracking the phase of interference
fringes in an interferometer is determined by the number of photons
arriving per coherence time (where the coherence time is again the RMS
time taken for the fringe phase to change by one radian). The photon
count entering the spatial filter per coherence time was calculated
for each aperture diameter and each level of AO correction, and these
are plotted in Fig.~\ref{fig:flux}. The drop in the photon counts for
large aperture sizes and partial AO correction is largely due to the
increased phase jitter and decreased coherence timescale for large
aperture sizes (see Figs.~\ref{fig:phase_jitter} and
\ref{fig:timescales}). For each level of AO correction, the aperture
diameter which provides the optimum signal-to-noise is marginally
larger than $D_{j}$ from Table~\ref{table:ao_list}. The Strehl ratio
provided by the AO system is $\sim~30\%$ at the optimum aperture size
in each case -- for simulations with Strehl ratios lower than this,
decreasing the aperture size would increase the photon count through
the spatial filter per coherence time if the AO system was still
matched to the aperture diameter.

Real AO systems do not perfectly compensate each Zernike mode
(particularly for faint off-axis reference stars), so additional
simulations were undertaken to determine the effect of noise in the
Zernike mode compensation on the RMS phase jitter. Every AO system has
different noise properties, and the calculation of the likely noise
spectrum from a real AO system is extremely complex. To keep the
simulations as simple and general as possible, the \emph{noise} was
simply modelled as being a small fraction of the atmospheric
corrugations. For example, in one simulation each of the corrected
Zernike modes was reduced to $10\%$ of the original amplitude, while
uncorrected modes were not adjusted. $44$ Zernike modes after piston
were corrected in all of these simulations, each mode being reduced to
between $5\%$ and $50\%$ of its original amplitude, depending on the
simulation. Aperture sizes from $4r_{0}$ to $20r_{0}$ were
investigated. Although the simulations presented here will not be
quantitatively accurate for any particular interferometer, they will
provide a qualitative description of what happens at all AO corrected
interferometers.

In all of the simulations the RMS phase jitter increased
dramatically whenever the Strehl ratio was $\mathrel{\hbox to
0pt{\lower 3pt\hbox{$\mathchar"218$}} \raise
2.0pt\hbox{$\mathchar"13C$}\hss}~30\%$ (corresponding to
$\mathrel{\hbox to 0pt{\lower 3pt\hbox{$\mathchar"218$}} \raise
2.0pt\hbox{$\mathchar"13E$}\hss}~1$~rad RMS wavefront error in the
aperture plane). In real AO systems the temporal properties of the
noise in the correction of Zernike modes is dependent on the bandwidth
of the AO system. When the Strehl ratio falls below $30\%$ we can
expect the coherence timescale for the fringes in the interferometer
to be determined by the temporal properties of the AO system, even if
the AO system does not introduce fluctuations in the aperture plane
piston mode.

\section{Conclusions}
\label{sect:conc}

Numerical simulations of atmospheric turbulence and AO wavefront
correction were performed to investigate the timescale for fringe
motion in optical interferometers with spatial filters. The fringe
motion was found to depend strongly on both the aperture diameter and
the level of AO correction used. In all of the simulations the
coherence timescale for interference fringes was found to decrease
dramatically when the Strehl ratio provided by the AO correction was
$\mathrel{\hbox to 0pt{\lower 3pt\hbox{$\mathchar"218$}} \raise
2.0pt\hbox{$\mathchar"13C$}\hss} 30\%$. For AO systems which give
perfect compensation of a limited number of Zernike modes, the optimum
aperture size for fringe phase tracking was calculated and found to be
dependent on the number of modes corrected. For AO systems which
provide noisy compensation of Zernike modes (but are perfectly
piston-neutral), the noise properties of the AO system were found to
determine the coherence timescale of the interference fringes when the
Strehl ratio is $\mathrel{\hbox to 0pt{\lower
3pt\hbox{$\mathchar"218$}} \raise 2.0pt\hbox{$\mathchar"13C$}\hss}
30\%$. These results highlight the need for time-resolved modelling of
the wavefront corrugations and fringe phase in existing AO-corrected
interferometers in order to fully understand the jitter seen in the
phase of the interference fringes.




\end{document}